\documentclass[prb,preprintnumbers,amsmath,amssymb]{revtex4}

\usepackage{graphicx}
\usepackage{dcolumn}
\usepackage{bm}

\begin{document}

\title{\bf Structural properties and enthalpy of formation of magnesium hydride from 
quantum Monte Carlo calculations}

\author{M. Pozzo$^1$}
\author{D. Alf\`{e}$^{1,2,3,4}$}%
\email{d.alfe@ucl.ac.uk}
\affiliation{
$^1$Department of Earth Sciences, University College London, Gower
Street, London WC1E 6BT, United Kingdom \\
$^2$Department of Physics and Astronomy, University College London, Gower
Street, London WC1E 6BT, United Kingdom \\
$^3$London Centre for Nanotechnology, University College London, 17-19 Gordon
Street, London WC1H 0AH, United Kingdom \\
$^4$Materials Simulation Laboratory, University College London, Gower Street, 
London WC1E 6 BT, United Kingdom\\}%

\date{\today}

\begin{abstract}

We have used diffusion Monte Carlo (DMC) calculations to study the
structural properties of magnesium hydride (MgH$_2$), including the
pressure-volume equation of state, the cohesive energy and the
enthalpy of formation from magnesium bulk and hydrogen gas.  The
calculations employ pseudopotentials and B-spline basis sets to expand
the single particle orbitals used to construct the trial
wavefunctions. Extensive tests on system size, time step, and other
sources of errors, performed on periodically repeated systems of up to
1050 atoms, show that all these errors together can be reduced to
below 10 meV per formula unit.  We find excellent agreement with the
experiments for the equilibrium volume of both the Mg and the MgH$_2$
crystals. The cohesive energy of the Mg crystal is found to be 1.51(1)
eV, and agrees perfectly with the experimental value of 1.51 eV. The
enthalpy of formation of MgH$_2$ from Mg bulk and H$_2$ gas is found
to be $0.85 \pm 0.01$ eV/formula unit, or $82 \pm 1$~kJ/mole, which is
off the experimental one of $76.1 \pm 1$~kJ/mole only by
$6$~kJ/mole. This shows that DMC can almost achieve chemical accuracy
(1 kcal/mole) on this system. Density functional theory errors are
shown to be much larger, and depend strongly on the functional
employed.

\end{abstract}

\maketitle

\section{Introduction}

The energetics of metal hydrides has recently become an issue of large
scientific and technological interest, mainly because of the revived
interest in these materials as potential hydrogen storage
media~\cite{schlapzutt01}. Magnesium hydride (MgH$_2$) is a
particularly interesting material, as it can store up to 7.6 \% of
hydrogen by weight, which is believed to be a large enough quantity
for mobile applications, provided that all the hydrogen in the
material can be made available when requested, of course. When heated
above $\sim 300~^o$C\cite{bogdanovic99} MgH$_2$ decomposes into Mg
bulk and H$_2$ gas, the reaction being endothermic with an enthalpy of
decomposition of 76 kJ/mole~\cite{yamaguchi94}. Conversely, MgH$_2$
can be synthesised by combining Mg bulk (usually in form of a powder of
micro-metre sized grains) and H$_2$ gas. The charging process can take
many hours, because of a large energy barrier to dissociate the H$_2$
molecule on the surface of magnesium~\cite{sprungplum91}.  As it
stands, MgH$_2$ is not considered to be useful for hydrogen storage
purposes, because of the high decomposition temperature (ideal
decomposition temperature should be in the range $20-100~^o$C), and the
slow kinetics of hydrogen intake.  A number of attempts are being made
to modify this material to improve its properties, including doping it
with traces of transition metals~\cite{liang99,shang04,hanada05},
which have been shown to be very effective at reducing the activation
energy for hydrogen dissociation ~\cite{du05,du06,du07,pozzo07}, and
also somewhat reduce the decomposition temperature of the
hydride~\cite{shang04}.

A number of theoretical calculations have been performed on magnesium
hydride and related systems (see for example ~\cite{yulam88} and
references therein; see also
~\cite{vajeeston02,shang04,song04,vansetten05,vegge05,du05,du06,du07,pozzo07}),
the most recent ones based on the implementation of quantum mechanics
known as density functional theory (DFT)~\cite{hohko64,kosha65}.
Although DFT can often be reliable at predicting trends in the
energetics of materials, it can be sometime in error when used to
obtain absolute energies.  In particular, as we show below, when
applied to the calculation of the enthalpy of formation of MgH$_2$,
the results are off by as much as 0.3 eV per formula unit, depending
on the functional employed, and cohesive energies can be wrong by over
0.5 eV.

Quantum Monte Carlo (QMC) techniques~\cite{foulkes01,umrigar93} are
believed to be one possible way to improve beyond density functional
theory. Since they are many order of magnitudes more computationally
demanding, the current database of properties of materials calculated
with QMC is still rather small, however, the increase in computer
power in the past few years is making now possible to perform
increasingly more numerous calculations on real systems, and
experience is being accumulated on the predictive power of this
technique.

Here we have used QMC to calculate the structural properties of the Mg
and MgH$_2$ crystals, together with their cohesive energies and the
enthalpy of formation of MgH$_2$ from Mg bulk and H$_2$ gas.  We find
excellent agreement with experiments for the structural properties of
the two solids, as well as the cohesive energy of the Mg solid. The
enthalpy of formation of MgH$_2$ is slightly overestimated, but the
error is of the order of 1 kcal/mole, showing that QMC on this system
can almost achieve chemical accuracy.

\section{Techniques}

\subsection{Density Functional theory calculations}

Density functional theory calculations have been performed with the
{\sc vasp} code~\cite{kresse96}. The interactions between the
electrons and the ionic cores was described using the projector
augmented method (PAW)~\cite{blochl94,kresse99} with the generalised
gradient approximations known as PBE~\cite{pbe}, PW91~\cite{pw91} or
the local density approximation (LDA). The Mg PAW potential has a
frozen Ne core and an outermost cutoff radius for the valence orbitals
of 1.06~\AA. The H PAW potential has a cutoff radius of
0.58~\AA. Single particle orbitals were expanded in plane waves with a
plane-wave cutoff of 270 eV, and a cutoff of 1600 eV was used for the
charge density. Such a large cutoff in the charge density (4 times
larger than the typical one used by default) is necessary to obtain
very accurate forces which are used to calculate the vibrational
properties of the crystals.  Calculations were performed by requiring
a self-consistency convergence on the total energy of $10^{-8}$~eV per
simulation cell. With these prescriptions convergence on the forces
was at worse equal to 0.2 meV/\AA, and one or two order of magnitudes
smaller for most atoms in the simulation cell. Brillouin Zone
integration was performed using {\bf k}-point sampling, with 18x18x12
and 10x10x15 Monkhorst-Pack~\cite{monkhorst76} grids on the Mg and
MgH$_2$ primitive cells respectively. With these densities of {\bf
k}-points the structural parameters are converged to better than 0.1
\%, and the total energies to better than 1 meV/primitive cell.

\subsection{Quantum Monte Carlo calculations}

Quantum Monte Carlo techniques have being extensively described
elsewhere~\cite{foulkes01,umrigar93}, so here we only report the main
technical details used in this work. Calculations have been performed using 
the CASINO code~\cite{casino}.  Diffusion Monte Carlo
calculations have been performed using trial wavefunctions of the
Slater-Jatrow type:
\begin{equation}
\Psi_T ( {\bf R} ) = D^\uparrow D^\downarrow e^{J} \; ,
\end{equation}  
where $D^\uparrow$ and $D^\downarrow$ are Slater determinants of up-
and down-spin single-electron orbitals, and $e^J$ is the so called
Jastrow factor, which is the exponential of a sum of one-body
(electron-nucleus), two-body (electron-electron) and three body
(electron-electron-nucleus) terms, which are parametrised functions of
electron-nucleus, electron-electron and electron-electron-nucleus
separations, and are designed to satisfy the cusp conditions. The
parameters in the Jastrow factor are varied to minimise the variance
of the local energy $E_L ( {\bf R} ) \equiv \Psi_T^{-1}({\bf
R})\hat{H} \Psi_T({\bf R})$.

Imaginary time evolution of the Schr\"odinger equation has been
performed with the usual short time approximation, and the locality
approximation~\cite{mitas91}. Time step errors have been carefully
analysed later in the paper.  Since the locality approximation
introduces an uncontrollable error with respect to which the DMC
energy is non-variational, we also tested the scheme of Casula~\cite{casula05}, 
which treats the non local part of the
pseudopotential in a consistent variational scheme.  We found that the
zero time step extrapolation of the energies in the two scheme
differed very little, which suggests that the errors in either case is
rather small~\cite{compar_la_casula}. However, we also found that the
time step error is much smaller in the locality approximation in this
particular case (this may not be true in general for other systems),
and therefore we decided to use the locality approximation throughout
the work which allowed us to work with a larger time step.

We used Dirac-Fock pseudopotentials (PP) for Mg and
H~\cite{trail05}. The Mg PP has a frozen Ne core and a core radius of
1.43~\AA, the H PP has a core radius of 0.26~\AA.  The single particle
orbitals have been obtained by DFT plane-wave (PW) calculations using
the LDA and a PW cutoff of 3400 eV, using the {\sc pwscf}
package~\cite{pwscf}. Such a large PW cutoff is due to the very small
H PP core radius, and was found to be necessary to reduce the variance
of the local energy as much as possible. We then exploited the
approximate equivalence between PW and B-splines~\cite{hernandez97} to
expand the single particle orbitals in a basis of B-spline, as
described in Ref.~\cite{alfe04}, using the natural B-spline grid
spacing given by $ a = \pi / G_{\rm max} $, where $G_{\rm max}$ is the
length of the largest vector employed in the PW calculations.

We used a diffusion Monte Carlo time step of 0.05 a.u., which was
found to result in errors of about 2 meV/f.u. (see below). With this
time step the acceptance ratios were 99.2 and 99.7 \% for the MgH$_2$
and Mg crystals respectively.  Total energies in the solids were
obtained by correcting the raw DMC data with DFT-LDA calculations
performed on the same cell size but a fully converged Brillouin zone
sampling, and then extrapolating these corrected DMC data to infinite
size (see below). The DMC calculations were performed using the Ewald
interaction to model electron-electron interactions.  The number of
walkers in the DMC simulations varied with the size of the systems,
and was never less than 1280.

\section{Results}

\begin{figure}
\centerline{
\includegraphics[width=3.5in]{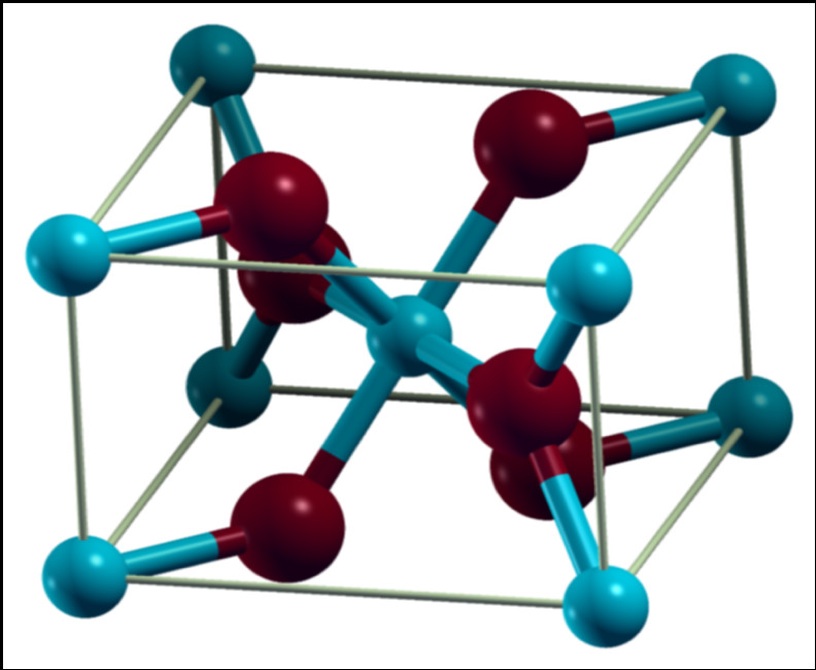}
}
\caption{\label{fig:mgh2_struct} (colour) Crystal structure of MgH$_2$ (see also text).
The Mg and H atoms are represented respectively by light blue and dark red colours.}
\end{figure}

Mg bulk has the hexagonal close packed structure, which is specified
by a lattice parameter $a$ and the ratio $c/a$ of the vertical axis to
one of the horizontal ones. The primitive cell contains two atoms, one
at the origin and the other at (1/3,2/3,0.5) in lattice vectors units.
The MgH$_2$ solid has a tetragonal structure of rutile type (see
Fig.~\ref{fig:mgh2_struct}), specified by a lattice parameter $a$ and the $c/a$
ratio. The primitive cell has two Mg atoms, one at the origin and the
other in the centre of the cell at (1/2,1/2,1/2) plus four hydrogen
atoms at ($\pm$x,$\pm$x,0) and (1/2$\pm$x,1/2$\mp$x,1/2). The exact
values of $c/a$ and x depend on pressure, and at ambient conditions
are found to be $c/a = 0.6687$ and x$=$0.304~\cite{bortz99}.

\subsection{Zero point energies and high temperature vibrational effects}

In order to compare the calculated structural parameters and cohesive
energies with the experimental ones we need to study the vibrational
properties of the crystals. This is because the experimental
parameters are usually determined at ambient conditions, and room
temperature thermal expansion for the Mg and MgH$_2$ solids is likely
to be significant.

We studied these vibrational properties within the quasi-harmonic
approximation, which far from the melting temperatures provides
accurate enough results for the thermal expansion of solids. This is
certainly the case for the Mg and MgH$_2$ solids at room
temperature. 

Phonons have been calculated using the {\sc phon} code~\cite{phon},
which implements the small displacement method~\cite{kresse95,alfe01}
to obtain the force constant matrix in crystals. The methods exploits
the linearity relation between the displacement of the atoms from
their equilibrium positions and the forces induced on all the atoms in
the crystal, which holds in the harmonic approximation for small
enough displacements.  The method is applied by constructing a
supercell which is a multiple of the primitive cell in the three
spacial directions, then the atoms in the primitive cell are displaced
by small amounts along three linearly independent directions and the
forces induced on all the atoms in the supercell are used to construct
the force constant matrix.  Symmetries can usually be used to reduce
the total number of displacements needed, and also to symmetrise the
force constant matrix~\cite{kresse95}. For bulk Mg, which has the
hexagonal closed packed crystal structure, only two displacements are
needed, one in the basal plane and one orthogonal to it (in fact, one
single off symmetry displacements would be sufficient, although this
would break the symmetry of the supercell and require a larger number
of {\bf k}-points in the DFT calculation of forces). MgH$_2$ has the
tetragonal structure of rutile TiO$_2$, with two Mg and four H atoms
in the primitive cell, and the total number of displacements needed in
this case is 4 (one could reduce the total number of displacements to
2 by sacrificing symmetries). If the supercell is large enough so that
the forces on the atoms sitting near the edges are small, then the
calculated force constant matrix becomes a good approximation of the
exact one.  Magnesium bulk is a metal, and convergence of the force
constant matrix with the size of the supercell is readily achieved: we
found that with cell containing 36 atoms (3x3x2) the ZPE is converged
to within 0.1 meV/atom (tested using supercells containing up to 150
atoms). However, MgH$_2$ is an insulator, and long range Coulomb
interactions make convergence slower. Nevertheless, we found that
already by using a cell containing 72 atoms (2x2x3 supercell) the ZPE
can be calculated with an accuracy of 0.5 meV/fu (tests used
supercells containing up to 576 atoms). All calculations were
performed with DFT-PBE.

Phonons calculated with the direct method described above may suffer
from inaccuracies due to the size of the displacements and/or
numerical noise in the calculated forces. To reduce the latter, one
would like to maximise the size of the displacements, but too large
displacements would cause departure from the harmonic regime. A
compromise between these two opposite requirements then needs to be
found, and this is usually achieved with displacement sizes of the
order of a fraction of a percent of the inter-atomic distances.
In order to test the size of the displacements we repeated the
calculations using displacements of 0.067~\AA, 0.04~\AA, 0.02~\AA, and
0.01~\AA, and we found that even with the largest displacement the ZPE
energy is converged to less than 0.2 meV/atom in Mg and 1 meV/fu in
MgH$_2$. We then decided to use displacements of 0.04~\AA. 

The fundamental vibrational frequency of the H$_2$ molecule has been
obtained by calculating the total energy of the H$_2$ molecule in a
large cubic box of size 13.5~\AA~for 5 different values of the H-H
distance, ranging from $R_0 - 0.0135$~\AA~to $R_0 + 0.0135$~\AA, where
$R_0 = 0.75$~\AA~is the calculated equilibrium distance with
DFT-PBE. The 5 energies have been fitted to a parabola, providing a
force constant of 33.35 eV/\AA$^2$ which corresponds to a stretching
vibrational frequency of 127 THz (only slightly lower that the
experimental value of 131.8 THz~\cite{bransden}), giving a ZPE of 0.263
eV.

\subsection{Density functional theory results}

Initially, we performed DFT calculations on the crystals with PBE,
PW91 and LDA.  Energy versus volume curves were fitted to a Birch-Murnaghan
equation of state~\cite{murna}, which provided equilibrium volumes and
bulk moduli. In the range of volumes considered, $c/a$'s do not change
very much from their zero pressure values, and the structural
parameters are essentially unchanged if $c/a$ is kept
fixed. Therefore, for simplicity we decided to fix $c/a$ to their
calculated zero pressure values of 1.621 and 0.6682 for Mg and MgH$_2$
respectively. The MgH$_2$ crystal has an additional degree of freedom,
which defines the position of the H atoms in the lattice. This has also
been optimised by fully relaxing the crystal at each different
volume. These relaxations are essential in the calculation of phonons,
because if the crystal is not in its ground state imaginary phonon
frequencies appear. However, as far as the energy is concerned, the
differences from calculations in which the H positions are kept at
their zero pressure equilibrium values are undetectable.

In Table~\ref{table1} we report the structural parameters of Mg and
MgH$_2$ calculated with the three density functionals, and we report
the results both at zero temperature (with and without ZPE) and at
room temperature.  Both Mg and MgH$_2$ are fairly soft materials, with
bulk moduli of the order of 40 and 50 GPa. Room temperature thermal
pressure are about 1 and 1.8 GPa for Mg and MgH$_2$ respectively,
which means volume thermal expansion is about 2\% and 3.5\% for the
two solids. This is significant, and cannot be ignored in a fair
comparison with the experimental data. We also report in the same
table the cohesive energies of the two solids.  The experimental
cohesive energy of MgH$_2$ can be estimated by combining the cohesive
energy of the Mg crystal (1.51 eV/atom), the dissociation energy of
the hydrogen molecule (4.48 eV/molecule) and the enthalpy of formation
of MgH$_2$ from Mg and H$_2$, whose value extrapolated at zero
temperature is $0.79 \pm 0.01$ eV/fu~\cite{yamaguchi94}, which
therefore give a result of $6.78 \pm 0.01$ eV/fu. By comparing the
calculated cohesive energies with the experimental ones it is clear
that the three functionals provide quite scattered results, with the
LDA doing better on MgH$2$ and PBE doing better on Mg. It is also
apparent that errors can be significant, of up 0.6 eV for PBE. This
error is well over 10 times a kcal/mole, which is the typical quantity
cited as {\it chemical accuracy}.

\begin{table}
\caption{\label{table1}Bulk properties (Volume/fu V$_0$ in \AA$^3$,
and bulk modulus k$_0$ in GPa) and cohesive energies (E$_{coh}$, in eV)
of Mg and MgH$_2$. Calculated properties are reported at zero
temperature with and without zero point energies (ZPE) and at the
temperatures at which the experimental data have been taken. Also
reported is the binding energy of the H$_2$ molecule }
\begin{ruledtabular}
\begin{tabular}{lccccc}
 & & \multicolumn{3}{c}{T$=0$ K} & {T\footnotemark[1]}  \\
\hline
 & & V$_0$, k$_0$ &  V$_0$, k$_0$ & E$_{coh}$ & V$_0$, k$_0$ \\
 & & (no ZPE) & (with ZPE) & (with ZPE) & \\
\hline 
Mg & & & & & \\
 & LDA & 21.59, 40.6 & 21.80, 39.3  & -1.74 & 22.14, 36.4 \\
 & PBE & 22.86, 36.5 & 23.08, 35.9 & -1.47 & 23.47, 34.0 \\
 & PW91 & 22.86, 36.4 & 23.10, 35.2 & -1.45 & 23.50, 32.6 \\
 & [Exp.] & & & [-1.51]\footnotemark[2] & [23.24\footnotemark[3], $36.8 \pm 3.0$ \footnotemark[4] \\
 & DMC & $22.96 \pm 0.05$, $35.5 \pm 1.2$ & $23.19 \pm 0.05$, $34.4 \pm 1.4$ & $-1.51 \pm 0.01$ & 
$23.61\pm 0.04$, $31.2 \pm 2.4$ \\ \\

MgH$_2$ & & & & & \\
 & LDA & 29.36, 55.5 & 30.32, 51.4 & -7.16 & 30.36, 49.9 \\
 & PBE & 30.84, 51.1 & 31.92, 45.8 & -6.17 & 32.03, 43.5 \\
 & PW91 & 30.72, 51.5 & 31.79, 46.4 & -6.27 & 31.89, 43.9 \\
 & [Exp.] & & & [$-6.78\pm 0.01$\footnotemark[5]\,] & [30.49\footnotemark[6], -- ] \\
 & DMC & $29.48 \pm 0.03$, $58.6 \pm 3.6$ & $30.53\pm 0.05$, $42.0 \pm 1.5$ & $-6.84 \pm 0.01$ & $30.58\pm  0.06$, $39.5 \pm 1.7$ \\ \\
H$_2$  & & & & & \\
 & LDA &             &            &  -4.59 & \\
 & PBE &             &            &  -4.23 & \\
 & PW91 &             &           &  -4.25 & \\
 & [Exp] &             &          &  [-4.48\footnotemark[7]\,] & \\
 & DMC &             &            &  $-4.484\pm 0.002$ & 
\footnotetext[1]{T$=298$ K for Mg, T$=260$ K for MgH$_2$.}
\footnotetext[2]{Ref.~\onlinecite{kittel}.}
\footnotetext[3]{Ref.~\onlinecite{webofel}.}
\footnotetext[4]{Ref.~\onlinecite{errandonea03}.}
\footnotetext[5]{Ref.~\onlinecite{yamaguchi94}.}
\footnotetext[6]{Ref.~\onlinecite{bortz99}.}
\footnotetext[7]{Ref.~\onlinecite{bransden}.}
\end{tabular}
\end{ruledtabular}
\end{table}

\subsection{Diffusion Monte Carlo results}

\subsubsection{Time step tests}

The dependence of the DMC energy on time step in the MgH$_2$ crystal
was studied by repeating simulations with a 2x2x3 supercell (72 atoms)
at time steps ranging from 0.005 to 0.15 a.u.. Calculations were
performed at the volume of 30.835~\AA$^3$/fu, and using the $A$ point
(0.5,0.5,0.5) which is at one corner of the Brillouin zone. For the Mg
crystal we used a 3x3x2 supercell (36 atoms), a volume of
22.785~\AA$^3$/atom and the H point (0.5,0.5,0.5), also at one corner
of the Brillouin zone.

\begin{figure}
\centerline{
\includegraphics[width=3.5in]{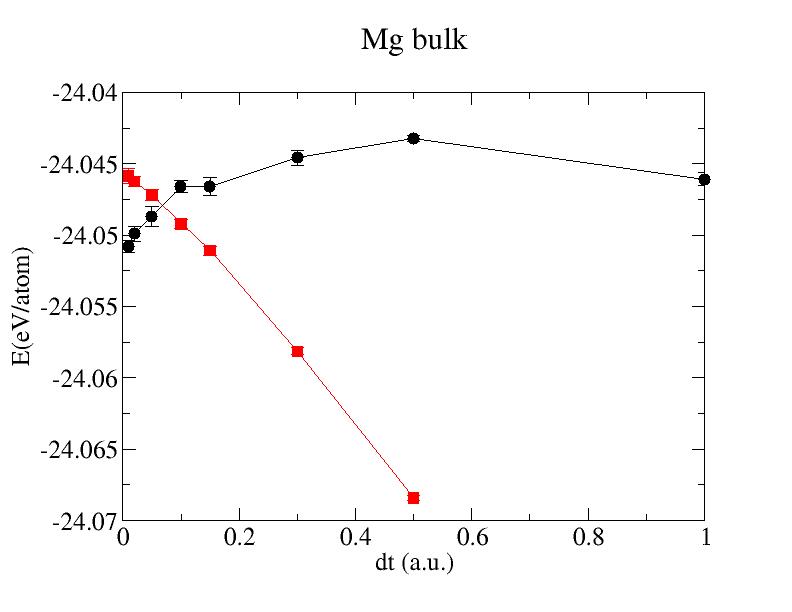}
\includegraphics[width=3.5in]{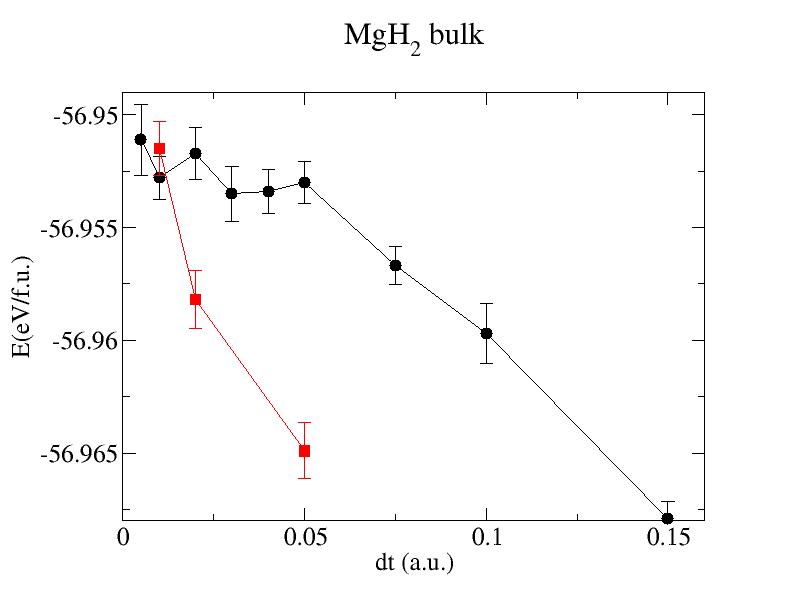}
}
\caption{Diffusion Monte Carlo energies for Mg bulk (left panel)
and MgH$_2$ bulk (right panel) as function of time step. Dots
and squares correspond to calculations performed with the locality
approximation and with the scheme proposed by Casula~\protect\cite{casula05} 
respectively.}
\label{fig:ts}
\end{figure}

Results of total energy/fu for MgH$_2$ and total energy/atom for Mg
are displayed in Fig.~\ref{fig:ts}, from which it is evident that
using a time step of 0.05 a.u. time step errors are well below 5
meV/fu. In Fig.~\ref{fig:ts} we also display the
results obtained with the scheme proposed by Casula~\cite{casula05}, and we 
observe that for short enough time steps
the two sets of energies are very close, and extrapolate to roughly
the same value in the limit of zero time step (to less than 5 meV/fu). 
As mentioned earlier, this suggests that the error introduced with either scheme is very
small. However, the locality approximation results in a much weaker
dependence of the DMC energy on time step and this is what we used
because it allowed us to work with much larger time steps.
We note that for the Mg crystal the time step error is much smaller, which 
in principle would allow us to work with larger time steps,
however, for consistency, we used the same time step of 0.05 a.u. also for the Mg crystal.

\begin{figure}
\centerline{
\includegraphics[width=3.5in]{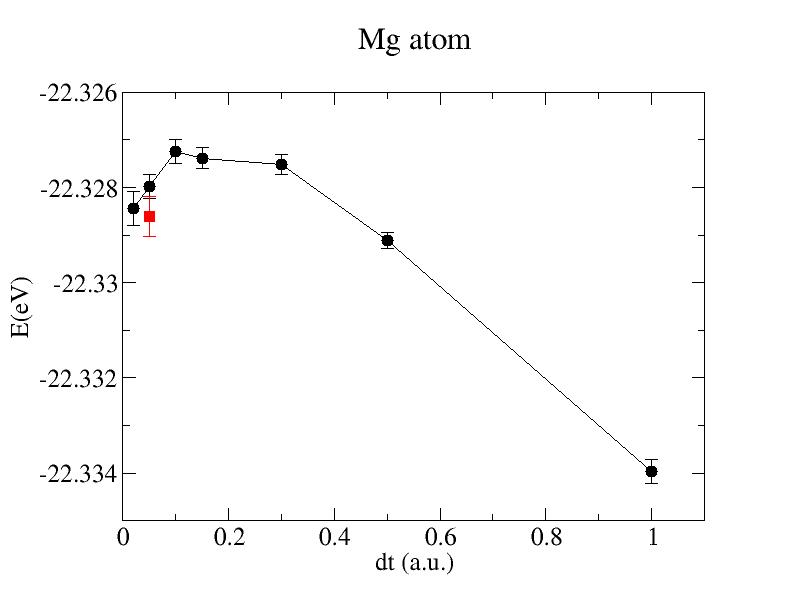}
\includegraphics[width=3.5in]{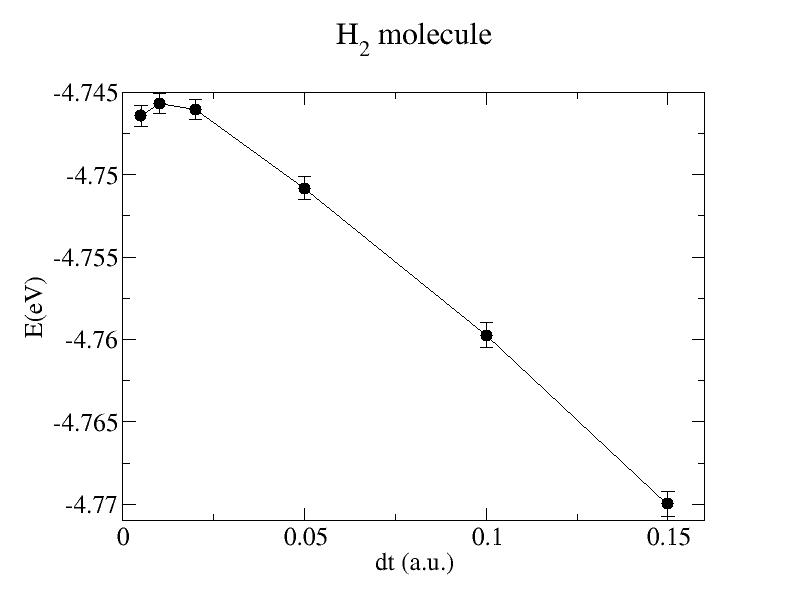}
}
\caption{Dots: diffusion Monte Carlo total energy for the Mg atom
(left panel) and binding energy of the H$_2$ molecule (right panel) as
function of time step. Calculations have been performed with the
locality approximation. Square: calculation performed with the scheme
proposed by Casula~\protect\cite{casula05}.}
\label{fig:tsatom}
\end{figure}

To calculate the total energies of the Mg atom and the H$_2$ molecule
we used trial wavefunctions obtained from plane wave calculations in
which the Mg atom or the H$_2$ molecule was placed at the centre of a
large cubic box with a side of 13.5~\AA. The DMC calculations were
then performed using B-splines and no periodically boundary
conditions.  We display in Fig.~\ref{fig:tsatom} the DMC energies as
function of time step, from which we can obtain very accurate zero
time step values. In the case of Mg we also performed one calculation
with the scheme of Casula~\cite{casula05}, which gave essentially
the same energy. For the H$_2$ molecule we display the binding energy
calculated at the equilibrium distance of 0.75~\AA, obtained
by subtracting from the energy of the molecule twice the energy of the
H atom, which is calculated to be 13.60635(5) eV. Both the energies of
the H atom and the H molecule are in excellent agreement with the
experimental data.

\subsubsection{The Mg crystal}

\begin{figure}
\centerline{
\includegraphics[width=3.5in]{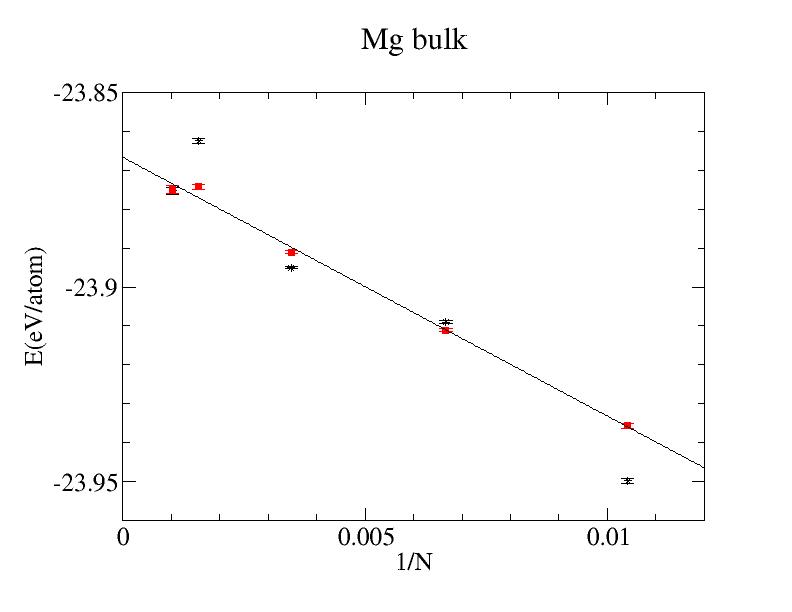}
}
\caption{Diffusion Monte Carlo total energy for the Mg crystal as
function of $1/N$, where N is the number of particles in the
simulation cell. Stars and squares correspond to raw and DFT corrected
(see text) results, solid line is a linear least square fit to the DFT
corrected results. }
\label{fig:sizemg}
\end{figure}

In the Mg crystal we studied the dependence of the DMC energy on the
size of the simulation cell by repeating the calculations with 4x4x3,
5x5x3, 6x6x4, 8x8x5 and 9x9x6 supercells, containing 96, 150, 288, 640
and 972 atoms respectively. Results are displayed in
Fig.~\ref{fig:sizemg}, where we show the total energies/atom $E_N$ as
function of $1/N$, with N the number of atoms in the simulation
cell. On the same graph we also show the energies $E_N^c = E_N + \left
[ E_\infty^{DFT} - E_N^{DFT} \right ]$, where $E_\infty^{DFT}$ are the
DFT energies calculated with fully converged {\bf k}-point sampling,
and $E_N^{DFT}$ are the DFT energies calculated with {\bf k}-point
samplings corresponding to the $N$-atom cells used in the DMC
calculations. It is clear that the raw DMC energies $E_N$ are quite
scattered and somewhat difficult to extrapolate to infinite size. This
is due to the metallic nature of Mg. However, the DFT corrected
energies $E_N^c$ are much better behaved, with data fitting quite well
onto a straight line, which makes it possible to extrapolate to infinite
size. In particular, we note that with no loss of accuracy we can also
use only the calculations with the 4x4x3, 5x5x3 and 6x6x4 supercells
to extrapolate to essentially the same infinite size value.

\begin{figure}
\centerline{
\includegraphics[width=3.5in]{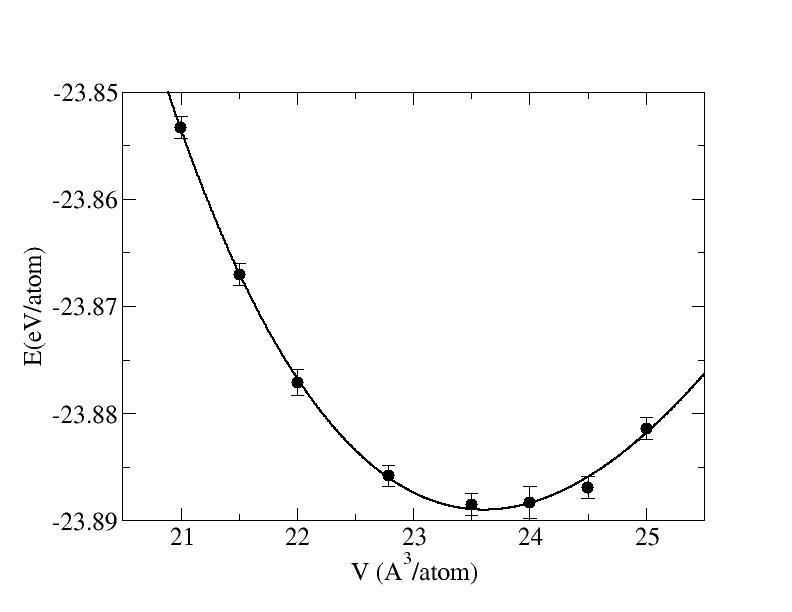}
}
\caption{Diffusion Monte Carlo free energies at 298~K for the Mg
crystal as function of volume $V$. Dots correspond to DMC
calculations extrapolated to infinite size, and include vibrational
free energies calculated with DFT-PBE. Solid line is a least squares
fit to a Birch-Murnaghan equation of state. }
\label{fig:mg}
\end{figure}

The calculations with these three supercell sizes were then repeated
at 8 different volumes, between 21 and 25~\AA$^3$/atom. At each volume
the DFT corrected DMC results were extrapolated to infinite size and
the results were fitted to a Birch-Murnaghan equation of state to
obtain the structural parameters. We performed the fit by weighing
each energy point $E_i$ point with $1/\sigma_i^2$, where $\sigma_i$ is
the standard error on $E_i$.  We report in Table~\ref{table1} the results
obtained both at zero temperature (with and without zero point energy)
and at room temperature. The latter are also shown in
Fig.~\ref{fig:mg}. The room temperature corrected DMC results slightly
overestimate the equilibrium volume, and also underestimate the bulk
modulus, but the calculated cohesive energy is in perfect
agreement with the experimental data.

\subsubsection{The MgH$_2$ crystal}

\begin{figure}
\centerline{
\includegraphics[width=3.5in]{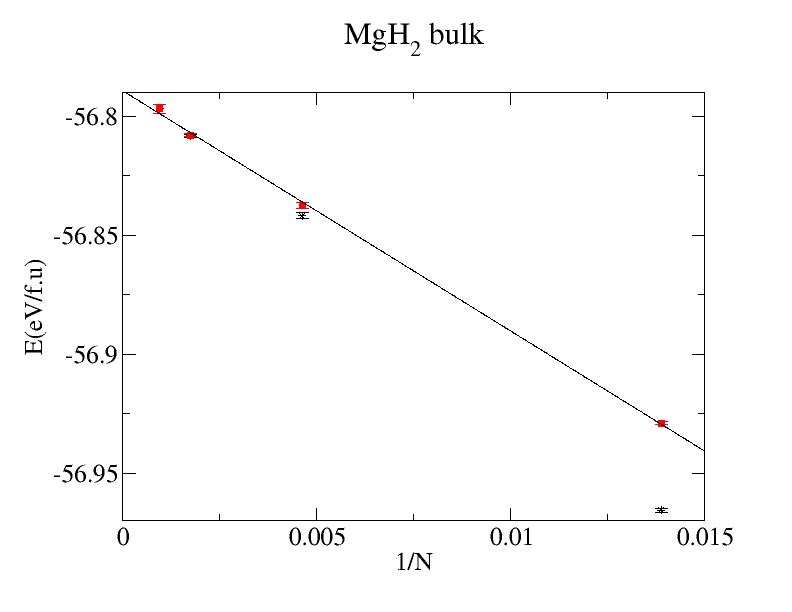}
}
\caption{Diffusion Monte Carlo total energy for the MgH$_2$ crystal as
function of $1/N$, where N is the number of particles in the
simulation cell. Stars and squares correspond to raw and DFT corrected
(see text) results, solid line is a linear least square fit to the DFT
corrected results.}
\label{fig:sizemgh2}
\end{figure}

For the MgH$_2$ crystal size effects were studied using 2x2x3, 3x3x4,
4x4x6 and 5x5x7 supercells, containing 72, 216, 576 and 1050 atoms
respectively. These tests were performed at the volume of
30~\AA$^3$/fu. The results for the four sizes studied are displayed in
Fig.~\ref{fig:sizemgh2}, where we show total energies/fu $E_N$ as
function of $1/N$, as well as the DFT corrected energies $E_N^c$. In
this case the DFT corrections are much smaller, which is not
surprising because of the large band gap in MgH$_2$.  A small
difference between the two sets of data can be observed for the
smallest sizes, but it is clear that they both fit very well onto
straight lines, which allows us to easily extrapolate the results to
infinite size. In fact, in this case the extrapolated results for the
two sets only differ by 5 meV/atom.

\begin{figure}
\centerline{
\includegraphics[width=3.5in]{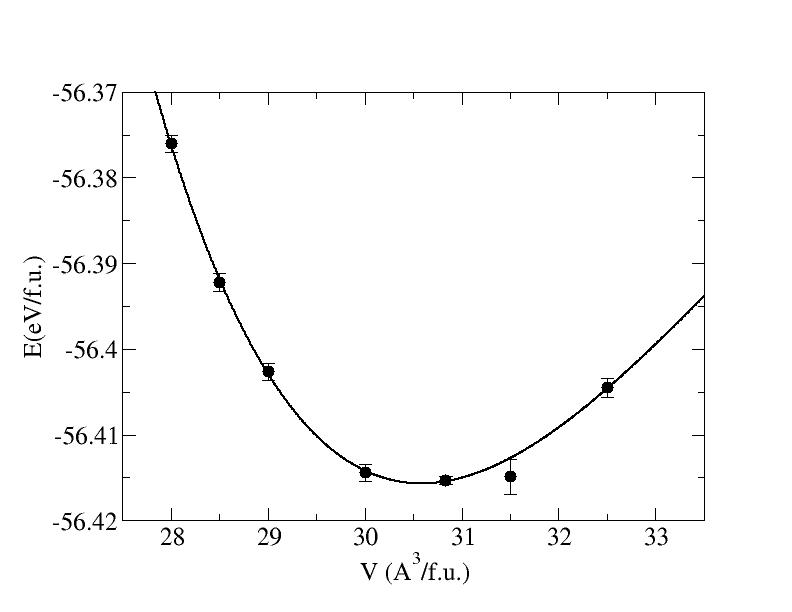}
}
\caption{Diffusion Monte Carlo free energies at 260~K for the MgH$_2$
crystal as function of volume $V$. Dots correspond to DMC
calculations extrapolated to infinite size, and include vibrational
free energies calculated with DFT-PBE.  Solid line is a least squares
fit to a Birch-Murnaghan equation of state. }
\label{fig:mgh2}
\end{figure}

The calculations were repeated at 7 different volumes between 28 and
32.5\AA$^3$/fu, and the DFT corrected DMC results were then fitted
with a Birch-Murnaghan equation of state to obtain the structural
parameters. Also in this case we used the inverse of the variances to
weigh each point in the fit. We report in Table~\ref{table1} the results
obtained both at zero temperature (with and without zero point energy)
and at T = 260~K, which is the temperature at which the experimental
data are reported~\cite{bortz99}. The high temperature results are
also shown in Fig.~\ref{fig:mgh2}. It is clear that once thermal
effects are added onto the calculations the agreement with the
experimental equilibrium volume is in almost perfect agreement. The
cohesive energy is slightly overestimated, but the error is only 0.06
eV, i.e. of the order of chemical accuracy.

\subsubsection{Enthalpy of formation of MgH$_2$}

We can now calculate the enthalpy of formation of MgH$_2$ from Mg bulk
and H$_2$ in the gas phase by adding the cohesive energies of the
MgH$_2$ and Mg crystals to the binding energy of the H$_2$ molecule.
We obtain enthalpy of formations of 0.82, 0.47 and 0.57 eV/fu with
LDA, PBE and PW91 respectively, and with DMC we obtain the value $0.85
\pm 0.01$~eV/fu. The LDA value is very accurate, but this is the
result of large cancellations of errors in the cohesive energies of
the crystals and the binding energy of the H$_2$ molecule. The DMC
result is only 0.06 eV higher than the experimental value of $0.79 \pm
0.01$ eV/fu, however in this case both the cohesive energies of the
crystals and the binding energy of the H$_2$ molecule are very
accurate.

\section{Conclusions}

We pointed out in this work the difficulty of using density functional
theory to calculate the enthalpy of formation of MgH$_2$ with high
accuracy. We studied the effect of three different
exchange-correlation functionals, PW91, PBE and LDA, and found that
although the GGA ones appear to work better on the Mg solid, the LDA
gives better results on the MgH$_2$ solid. It turns out, therefore,
that is difficult to get a good DFT value for the enthalpy of
formation of MgH$_2$: the two GGA functionals give an enthalpy of
formation in error of more than 0.2 and 0.3 eV/fu respectively. The
LDA is the functional that does best, but for the wrong reason,
because the cohesive energies of the crystals and the binding energies
of the molecule are wrong by up to 0.4 eV/fu, and the enthalpy of
formation is accurate only because of large cancellation of errors.

Diffusion Monte Carlo appears to deliver much better accuracy in
general. We have shown that the DMC equilibrium volumes of MgH$_2$
agrees perfectly with the experimental one, once high temperature
thermal expansion is included in the calculations, and the equilibrium
volume of Mg is only slightly overestimated. The cohesive energy of Mg
is also predicted in perfect agreement with the experimental datum,
and so is the binding energy of the H$_2$ molecule. A small error is
present in the cohesive energy of the MgH$_2$ crystal, which
determines the small inaccuracy in the enthalpy of formation, for
which we find a DMC value of $0.85 \pm 0.01$~eV/fu. However, this is
only 0.06 higher than the accepted experimental one of $0.79 \pm
0.01$~eV/fu, or $76.1 \pm 1$~kJ/mole. This result is not very far from
the LDA value, but with the important difference that now all three
terms that enter the enthalpy of formation are calculated accurately,
and we don't rely on fortunate cancellation of errors. 

Although the DMC error  is slightly larger than 1 kcal/mole, and therefore 
we cannot claim {\it chemical accuracy}, we are not far from it, and therefore 
we argue that quantum Monte Carlo techniques have useful predictive power 
in the search of metal hydrides with workable decomposition temperatures.

\section*{Acknowledgments}
The computations were performed on the HPCx service, using allocations
of time from NERC, and on HECToR. Calculations have also been performed on the LCN
cluster at University College London.  
This work was conducted as part of a EURYI scheme award as provided by
EPSRC (see www.esf.org/euryi).

\begin{thebibliography}{99}

\bibitem{schlapzutt01} L. Schlapbach and A. Z\"{u}ttel, Nature
(London) {\bf 414}, 353 (2001).

\bibitem{bogdanovic99} B. Bogdanovic, K. Bohmhammel, B. Christ, A. Reiser, K. Schlichte, 
R. Vehlen and U. Wolf, J. Alloys Comp. {\bf 282}, 84 (1999).

\bibitem{yamaguchi94} M. Yamaguchi and E. Akiba, in {\it Material Science and Technology}, 
vol. 3B, edited by R. W. Cahn, P. Haasen and E. J. Kramer (New York: VCH, 1994), p. 333. 

\bibitem{sprungplum91} P. T. Sprunger and E. W. Plummer, Chem. Phys. Lett. {\bf 187}, 559 
(1991).  

\bibitem{liang99} G. Liang, J. Huot, S. Boily, A. Van Neste and R. Schulz, J. Alloys 
Compd. {\bf 292}, 247 (1999). 

\bibitem{shang04} C. X. Shang, M. Bououdina, Y. Song and Z. X. Guo, Int. J. Hydrogen 
Energy {\bf 29}, 73 (2004).

\bibitem{hanada05} N. Hanada, T. Ichigawa and H. Fujii, J. Phys. Chem. B  {\bf 109}, 
7188 (2005).

\bibitem{du05} A. J. Du, S. C. Smith, X. D. Yao, and G. Q. Lu, J. Phys. Chem. B 
{\bf 109}, 18037 (2005).

\bibitem{du06} A. J. Du, S. C. Smith, X. D. Yao, and G. Q. Lu, J. Phys. Chem. B 
{\bf 110}, 21747 (2006).

\bibitem{du07} A. J. Du, S. C. Smith, X. D. Yao, and G. Q. Lu, J. Am. Chem. Soc. 
{\bf 129}, 10201 (2007).

\bibitem{pozzo07} M. Pozzo, D. Alf\`e, A. Amieiro, S. French and A. Pratt, 
unpublished.

\bibitem{yulam88} R. Yu and P. K. Lam, Phys. Rev. B {\bf 37}, 8730 (1988).

\bibitem{vajeeston02} P. Vajeeston, P. Ravindran, A. Kjekshus, and H. Fjellv\r{a}g,
Phys. Rev. Lett. {\bf 89}, 175506 (2002). 

\bibitem{song04} Y. Song, Z. X. Guo and R. Yang, Phys. Rev. B {\bf 69}, 094205 (2004).

\bibitem{vansetten05} M. J. van Setten, G. A. de Wijs, V. A. Popa, and G. Brocks, 
Phys. Rev. B {\bf 72}, 073107 (2005).

\bibitem{vegge05} T. Vegge, L. S. Hedegaard-Jensen, J. Bonde, T. R. Munter and 
J. K. N{\o}rskov, J. Alloys Compd. {\bf 386}, 1 (2005).

\bibitem{hohko64} P. Hohenberg, W. Kohn, Phys. Rev. B {\bf 136}, 864 (1964).

\bibitem{kosha65} W. Kohn and L.J. Sham. Phys. Rev. A {\bf 140}, 1133 (1965). 

\bibitem{foulkes01} W. M. C. Foulkes, L. Mit\'a\v{s}, R. J. Needs, and
G. Rajagopal, Rev. Mod. Phys. {\bf 73}, 33 (2001).

\bibitem{umrigar93} C. J. Umrigar, M. P. Nightingale, K. J. Runge,
J. Chem. Phys., {\bf 99}, 2865 (1993).

\bibitem{casino} R.J. Needs, M.D. Towler, N.D. Drummond and P. Lopez Rios, CASINO 
version 2.1 User Manual, University of Cambridge, Cambridge (2007).

\bibitem{kresse96} G. Kresse and J. Furthmuller,
Comput. Mater. Sci. {\bf 6} (1996) 15; Phys. Rev. B {\bf 54}, 11169 (1996).

\bibitem{blochl94} P. E. Bl\"{o}chl, Phys. Rev. B, {\bf 50}, 17953 (1994).

\bibitem{kresse99} G. Kresse and D. Joubert, Phys. Rev. B, {\bf 59},
1758 (1999).

\bibitem{pbe} J.P. Perdew, K. Burke and M. Ernzerhof. Phys. Rev. Lett. {\bf 77},
3865 (1996). 

\bibitem{pw91} J.P. Perdew and Y. Wang. Phys. Rev. B {\bf 45}, 13244 (1992).

\bibitem{monkhorst76} H. J. Monkhorst and J. D. Pack, Phys. Rev. B
{\bf 13}, 5188 (1976).

\bibitem{mitas91} L. Mit\'a\v{s}, E. L. Shirley, D. M. Ceperley,
J. Chem. Phys. {\bf 95}, 3467 (1991).

\bibitem{casula05} M. Casula, 
Phys. Rev. B {\bf 74}, 161102 (2006).

\bibitem{compar_la_casula} Of course, it is possible that the Casula
scheme induces a large positive error in the DMC energy, and that by
chance the error induced by the locality approximation is also
positive and of the same size. However, since the two scheme are
completely different, we believe that this is unlikely.

\bibitem{trail05} J. R. Trail and R. J. Needs, J. Chem. Phys. {\bf
122}, 014112 (2005); ibid J. Chem. Phys. {\bf 122}, 174109 (2005), 
see also www.tcm.phy.cam.ac.uk/$\sim$mdt26/casino2$\_$pseudopotentials.html.

\bibitem{pwscf} S. Baroni, A. Dal Corso, S. de Gironcoli, and
P. Giannozzi, http://www.pwscf.org.

\bibitem{hernandez97}
E. Hern\'{a}ndez, M. J. Gillan and C. M. Goringe, Phys. Rev. B
{\bf 55},  13485 (1997).

\bibitem{alfe04} D. Alf\`e and M. J. Gillan, Phys. Rev B, {\bf 70}, 161101(R)
(2004).
\bibitem{bortz99} M. Bortz, B. Bertheville, G. B\"ottger, K. Yvon,
J. Alloys Compd. {\bf 287}, L4 (1999)

\bibitem{phon} D. Alf\`e, (1998). Program available at
http://chianti.geol.ucl.ac.uk/$\sim$dario;
D. Alf\`e,  G. D. Price, M. J. Gillan, Phys. Rev. B, {\bf 64}, 04512316 (2001).

\bibitem{kresse95} G. Kresse, J. Furthm\"{u}ller and J. Hafner,
Europhys. Lett. {\bf 32}, 729 (1995).

\bibitem{alfe01} D. Alf\`e, G. D. Price, M. J. Gillan, Phys. Rev. B,
{\bf 64}, 045123 (2001).

\bibitem{bransden} B. H. Bransden and C. J. Joachin, {\it Physics of Atoms
and Molecules} (Wiley, New York, 1983).

\bibitem{murna} F. Birch, Phys. Rev. {\bf 71}, 809 (1947).
\bibitem{kittel} C. Kittel, {\it Introduction to Solid State Physics}, 7th ed. (Wiley, 
New York, 1996).
\bibitem{webofel} http://www.webelements.com
\bibitem{errandonea03} D. Errandonea, Y. Meng, D. H\"{a}usermann, and T. Uchida, 
J. Phys.: Condens. Matter {\bf 15}, 1277 (2003).

\end{thebibliography}
\end{document}